\documentclass{PoS}

\PoS{PoS(HEP2005)016}

\usepackage{verbatim}
\usepackage{graphicx}
\usepackage{url}
\usepackage{amsfonts}
\usepackage{amsmath,amssymb}

\newcommand{\lb}[0] { \left( }
\newcommand{\rb}[0] { \right) }

\newcommand{\beqs} { \begin{eqnarray} }
\newcommand{\eeqs} { \end{eqnarray} }
\newcommand{\bsub}{ \begin{subequations} }
\newcommand{\esub}{ \end{subequations} }

\renewcommand{\eqref}[1]{(\ref{#1})}
\newcommand{\mE} {\mathcal{E}}

\newcommand{\ergscm}[0] {\textrm{ erg s$^{-1}$ cm$^{-3}$}}

\newcommand{\erg}[0] {\textrm{ erg}}
\newcommand{\MeV}[0] {\textrm{ MeV}}
\newcommand{\Kel}[0] {\textrm{ K}}
\newcommand{\cm}[0] {\textrm{ cm}}

\newcommand{\EE}[2] {#1 \times 10^{#2}}

\title{Thermal neutrinos from hot GRB fireballs}
\ShortTitle{Thermal neutrinos from hot GRB fireballs}

\author{\speaker{Hylke B. J. Koers}\\
        NIKHEF \\
        PO Box 41882, 1009 DB Amsterdam, The Netherlands \\
        \emph{and} \\
        University of Amsterdam, 
        Amsterdam, The Netherlands
        \\
        E-mail: \email{hkoers@nikhef.nl}}

\author{Ralph A.M.J. Wijers\\
        Astronomical Institute `Anton Pannekoek' \\ Faculty of Science, University of Amsterdam \\
        Kruislaan 403, 1098 SJ Amsterdam, The Netherlands \\ \\ \emph{NIKHEF-2005-021} \\ \emph{astro-ph/0511071}}

\abstract{
We consider the physics of neutrinos in a fireball, i.e. a
tightly coupled plasma of photons, positrons and electrons.
Such a fireball is believed to form in the first stages of a gamma-ray 
burst. We assume the fireball is radiation-dominated and spherically 
symmetric. Energy considerations limit the allowed baryon density, from 
which it follows that the neutrino physics is dominated by leptonic 
processes. We find that, for quite general initial conditions, neutrinos 
start out in thermodynamic equilibrium with the fireball and follow the 
usual hydrodynamical evolution. As the fireball cools, the plasma becomes 
transparent to neutrinos which subsequently decouple from the plasma. 
Although a sizable fraction of the total energy is carried away, the 
detection possibility of these neutrino bursts is limited due to the 
isotropic outflow and the relatively low mean energy of approximately 60 
MeV.
}

\FullConference{International Europhysics Conference on High Energy Physics\\
        July 21st - 27th 2005\\
        Lisboa, Portugal}

\begin{document}

\section{Introduction}
The current paradigm on long gamma-ray bursts (GRBs) is that they are initiated by the violent
collapse of a massive star (see e.g. \cite{Paradijs2000}). In this process, a black hole -- accretion disk 
system is formed, which stores huge amounts of energy. 
Assuming that a fair fraction of the
gravitational energy is liberated into a volume contained within a few Schwarzschild radii, one expects the formation of a  fireball, a photon-electron-positron plasma with
a small baryonic load \cite{1978MNRAS.183..359C}. Radiation
pressure accelerates the fireball to high Lorentz factors. This process
converts thermal energy to kinetic energy of the 
baryons, which is again dissipated by shock acceleration in the optically thin region far away from the central object. This results in the observed gamma-ray bursts.

The physics of these photon-electron-positron fireballs was explored by Cavallo and Rees \cite{1978MNRAS.183..359C}.
With the cosmological distances and  higher energies presently believed to be involved, 
the fireball could be hot enough to contain neutrinos\footnote{We will generally use `neutrinos' for `neutrinos and antineutrinos'.}.
We investigate the role of neutrinos in the dynamical evolution of the fireball and 
the expected neutrino emission.

\section{The fireball}
\label{sect:general}
The parameter space of a generic fireball 
can be divided into three regions: region I where the fireball contains neutrinos of all flavors; region II where it contains only electron-neutrinos; and region III where all neutrinos are decoupled. 
In thermodynamic equilibrium, the energy density and temperature are related through $\mE/V =  g a  T^4$, 
where $a$ is the radiation constant and
$ g_{\textrm{I}} =  43/8$;
$ g_{\textrm{II}}=  29/8$;
$ g_{\textrm{III}} = 11/4$.
We assume a spherical configuration
and adopt the following reference values for the initial fireball energy, radius and temperature:
\beqs
\label{eq:refs}
\mE_*  =   10^{52} \erg \, , \qquad 
R_*  =   10^{6.5} \cm \, , \qquad
T_* = \EE{2.1}{11} \Kel  = 17.9 /k_B \MeV \, .
\eeqs
Typical number densities for particles in thermal equilibrium are of
the order $n \sim 10^{35}$ cm$^{-3}$. The nucleon\footnote{Because the temperature is higher than typical binding energies, all nuclei are dissociated 
into nucleons.} density is restricted by energy considerations: we require \mbox{1 TeV} per nucleon
for acceleration to sufficiently large Lorentz factors. We take this upper limit on the density as 
 the reference value:
$n_{B,*} =  \EE{5}{31} \cm^{-3}$ . 
Because of overall charge neutrality, the net electron
density \mbox{$\Delta n_e :=  n_{e^-} - n_{e^+}$} equals
the proton density. The low nucleon density then implies that
\mbox{$\Delta n_e \ll n_{e^-} + n_{e^+}$}, so that the electron chemical potential is very small:
\mbox{$\mu_e / (k_B T) \ll 1$}.

\section{Fireball neutrino physics}
\label{sect:neutrinos}
The neutrino physics in this environment is dominated by leptonic processes 
(for a more elaborate discussion on the various processes, see \cite{Koers2005}).
The dominant neutrino
production process is electron-positron pair annihilation
\mbox{$e^- + e^+ \to \nu +\bar{\nu}$}, with total emissivity
\cite{Dicus:1972yr, 1989ApJ...339..354I}
\beqs
Q_{\textrm{pair}} =  \EE{3.6}{33} \,  \lb T_{11} \rb^9 \ergscm 
\, , \quad
\textrm{where}
\quad T_{11} = T / 10^{11} \, \textrm{K}  \, . 
\eeqs
The neutrino mean free path (mfp) is set
by scattering off electrons and positrons \cite{1975ApJ...201..467T}:
\bsub
\label{eq:op:epscat}
\beqs
\lambda^{(\textrm{e})}  =  \EE{3.7}{6} \,  \lb T_{11} \rb^{-5} \cm \, , \qquad
\lambda^{(\mu,\tau)}  =  \EE{1.6}{7}  \, \lb T_{11} \rb^{-5}  \cm \, .
\eeqs
\esub
The mfps for neutrinos and
for antineutrinos are virtually equal because of the equal amounts of electrons
and positrons. We express the neutrino creation rate in terms of the parameter $\chi = t_c / t_e$, where $t_c= \mE / ( V Q )$ is the cooling timescale 
and  $t_e= R / c_s$ is the expansion timescale:
\beqs
\label{eq:phase:chi}
\chi = \EE{3.7}{-3} \, g^{9/4} \lb \mE_{52} \rb^{-5/4} \lb R_{6.5} \rb^{11/4} 
\, , \quad \mE_{52} = \mE / 10^{52} \, \textrm{erg}  
\, , \quad
R_{6.5} = R / 10^{6.5} \, \textrm{cm}  \, .
\eeqs
We used $c_s = c / \sqrt{3}$ for the sound speed in the plasma. For the reference values
of eq. \eqref{eq:refs}, we find
that $\chi_{\textrm{I}}  = 0.16$. This means that neutrinos are created  reasonably rapidly compared
to the expansion timescale.
The fireball's opacity to neutrinos is described in terms of the optical depth \mbox{$\tau = R / \lambda$}, where  $R$ is
the length scale and $\lambda$ is the mfp:
\beqs
\label{eq:phase:tau}
\tau^{(\textrm{e})}    =    54 \times  \lb \mE_{52} \rb^{5/4} \lb R_{6.5} \rb^{-11/4} \, , \qquad
\tau^{(\mu,\tau)}    =   7.4 \times  \lb \mE_{52} \rb^{5/4} \lb R_{6.5} \rb^{-11/4} \, .
\eeqs
For reference initial conditions, \mbox{$\tau^{(\textrm{e},\mu,\tau)} > 1 $} so that the fireball is opaque to neutrinos of all flavors.

\section{Evolution of the neutrino fireball}

\label{sect:evolution}
\begin{figure*}
\center
\includegraphics[height=5.2cm]{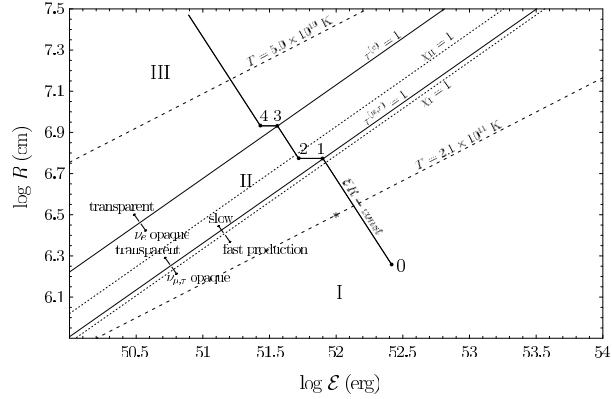} 
\caption{The parameter space of the fireball.  The solid lines
show the $\chi_{\textrm{I}}=1$,  $\chi_{\textrm{II}}=1$,  $\tau^{(\mu,\tau)}=1$ and $\tau^{(\textrm{e})}=1$ contours; the dotted lines are isotemperature curves. The $*$
denotes the reference point of eq. (2.1).}
\label{figure:phase1}
\end{figure*}
Figure \ref{figure:phase1} shows how the parameter space of the neutrino-fireball is divided in the regions I, II and III by
the $\tau^{(\mu,\tau)}=1$ and $\tau^{(\textrm{e})}=1$ contours. Also shown is
the  neutrino creation rate parameter $\chi$. 
In region I neutrinos are created rapidly
compared to the expansion
timescale while the fireball is neutrino-opaque, 
so that thermodynamic equilibrium will be established rapidly.

The evolution of the fireball is very similar to that of a neutrinoless fireball (see e.g. \cite{1990ApJ...365L..55S}).
The plasma expands by radiation pressure, converting radiative energy to kinetic energy of the baryons. We assume that the
expansion is adiabatic and reversible, so that
$ R T = \textrm{const} $ and $ \mathcal{E} R = \textrm{const}$
during the expansion (see ref. \cite{1990ApJ...365L..55S}; note that $\mE$ denotes the thermal energy). If one of the plasma components decouples, the temperature-radius relationship
still holds  but a fraction of the energy is lost at constant radius \cite{Koers2005}.

Consider the evolution of a fireball that starts with initial energy $\mE_0$ and size $R_0$
in region I. The trajectory is sketched in
figure \ref{figure:phase1}. 
Starting from the point denoted as `0' in figure \ref{figure:phase1}, the plasma expands
along a $\mathcal{E} R = \mathcal{E}_0 R_0$ line until it
reaches the $\tau^{(\mu,\tau)} = 1$ contour,
where the muon- and tau-neutrinos decouple from the plasma. 
These carry away $14/43 \simeq 33$\% of the
thermal energy that is then available (a fraction of the initial energy is already converted to kinetic energy of the nucleons). This moves the fireball from point 1 to point 2.
The electron-neutrinos
remain in  thermal equilibrium with the plasma, which expands along a $\mathcal{E} R = \mathcal{E}_2 R_2$ curve. When the plasma becomes transparent to electron-neutrinos at $\tau^{(\textrm{e})} =1$,
these carry away $7/29  \simeq 24$\%  of the energy (point 4).

\section{Neutrino emission}
Apart from the decoupling bursts, neutrinos are emitted continuously 
in regions where the creation rate is sufficiently high and the plasma is transparent to neutrinos. This continuous emission component is
smaller than the bursts and never causes dramatic energy losses.
The spectrum of the emitted neutrinos is thermal, with observed temperature roughly equal to the initial temperature of the plasma \cite{Goodman:1986az,Koers2005}. We find a total energy  and  mean energy of
\beqs
\label{E:nu:burst}
E_{(\nu \textrm{ tot})}  =  3.1 \times 10^{52} \, \textrm{erg} 
\lb \mE_{52}^{(0)} \, R^{(0)}_{6.5} \rb^{11/16} \, , \quad
\langle E_\nu \rangle  =   56 \MeV  \lb \mE_{52}^{(0)} \rb^{1/4} \lb R_{6.5}^{(0)} \rb^{-3/4} \, .
\eeqs
Because the fireball has not expanded much in between the two decoupling events (see figure \ref{figure:phase1}), the intrinsic
time spread is determined by the size of the fireball at the second
burst: $\Delta t  \sim 0.4 $  ms. Due to dispersion, the observed time spread is expected to be
around $\Delta t \sim 1$ ms. Because of the relatively low mean energy and isotropic outflow, detection
possibilities of such a neutrino source are limited to a few Mpc (see ref.  \cite{Koers2005} for a
discussion and references).

\section{Conclusions}
\label{sect:conclusions}
We have described the physics of neutrinos in a hot fireball plasma. For parameters that are thought likely for the initial fireballs in gamma-ray bursts, we find that the dominant neutrino processes are leptonic
and thus cooling is equally rapid in baryon-free
fireballs. A fireball that
starts in the neutrino-opaque region I creates neutrinos rapidly enough to establish thermodynamic equilibrium. The neutrinos follow the usual hydrodynamical
evolution of the fireball until decoupling. This results in a ms burst of $\sim$60 MeV neutrinos, which carry  away $\sim$30\% of the initial energy in the fireball. Although this is a sizable fraction, neutrino
cooling is never dramatic and the neutrinos will be difficult to detect.

\end{document}